\documentclass[iop]{emulateapj}

\newcommand{\sname}{W0204--0506}
\newcommand{\CIV}{C\,{\sc iv}\,$\lambda$1549\AA}
\newcommand{\HeII}{He\,{\sc ii}\,$\lambda$1640\AA}
\newcommand{\CIII}{C\,{\sc  iii}]\,$\lambda$1909\AA}
\newcommand{\civ}{C\,{\sc iv}}
\newcommand{\heii}{He\,{\sc ii}}
\newcommand{\ciii}{C\,{\sc  iii}]}
\newcommand{\cii}{[C\,{\sc  ii}]}

\begin{document}

\title{Hot Dust Obscured Galaxies with Excess Blue Light: Dual AGN or
  Single AGN Under Extreme Conditions?}

\author{R.J.~Assef\altaffilmark{1}, 
  D.J.~Walton\altaffilmark{2}, 
  M.~Brightman\altaffilmark{2}, 
  D.~Stern\altaffilmark{3}, 
  D.~Alexander\altaffilmark{4},
  F.~Bauer\altaffilmark{5,6,7}, 
  A.W.~Blain\altaffilmark{8},
  T.~Diaz-Santos\altaffilmark{1},
  P.R.M.~Eisenhardt\altaffilmark{3},
  S.L.~Finkelstein\altaffilmark{9},
  R.C.~Hickox\altaffilmark{10},
  C.-W.~Tsai\altaffilmark{3},
  J.W.~Wu\altaffilmark{11}
}

\altaffiltext{1}{N\'ucleo de Astronom\'ia de la Facultad de
  Ingenier\'ia, Universidad Diego Portales, Av. Ej\'ercito Libertador
  441, Santiago, Chile. E-mail:~{\tt{roberto.assef@mail.udp.cl}}}

\altaffiltext{2}{Space Radiation Laboratory, California Institute of
  Technology, Pasadena, CA 91125, USA}

\altaffiltext{3}{Jet Propulsion Laboratory, California Institute of
  Technology, 4800 Oak Grove Drive, Mail Stop 169-236, Pasadena, CA
  91109, USA.}

\altaffiltext{4}{Department of Physics, Durham University, Durham DH1
  3LE, UK.}

\altaffiltext{5}{Departamento de Astronom\'ia y Astrof\'isica,
  Pontificia Universidad Cat\'olica de Chile, Casilla 306, Santiago
  22, Chile}

\altaffiltext{6}{Space Science Institute, 4750 Walnut Street, Suite
  205, Boulder, CO 80301, USA}

\altaffiltext{7}{Millennium Institute of Astrophysics, MAS, Nuncio
  Monse\~{n}or S\'{o}tero Sanz 100, Providencia, Santiago de Chile}

\altaffiltext{8}{Physics \& Astronomy, University of Leicester, 1
  University Road, Leicester LE1 7RH, UK}

\altaffiltext{9}{The University of Texas at Austin, 2515 Speedway,
  Stop C1400, Austin, Texas 78712, USA.}

\altaffiltext{10}{Department of Physics and Astronomy, Dartmouth
  College, 6127 Wilder Laboratory, Hanover, NH 03755, USA}

\altaffiltext{11}{UCLA Astronomy, PO Box 951547, Los Angeles, CA
  90095-1547, USA}

\begin{abstract}
Hot Dust-Obscured Galaxies (Hot DOGs) are a population of
hyper-luminous infrared galaxies identified by the WISE mission from
their very red mid-IR colors, and characterized by hot dust
temperatures ($T>60~\rm K$). Several studies have shown clear evidence
that the IR emission in these objects is powered by a highly
dust-obscured AGN that shows close to Compton-thick absorption at
X-ray wavelengths. Thanks to the high AGN obscuration, the host galaxy
is easily observable, and has UV/optical colors usually consistent
with those of a normal galaxy. Here we discuss a sub-population of 8
Hot DOGs that show enhanced rest-frame UV/optical emission. We discuss
three scenarios that might explain the excess UV emission: (i)
unobscured light leaked from the AGN by reflection over the dust or by
partial coverage of the accretion disk; (ii) a second unobscured AGN
in the system; or (iii) a luminous young starburst. X-ray observations
can help discriminate between these scenarios. We study in detail the
blue excess Hot DOG WISE J020446.13--050640.8, which was
serendipitously observed by {\it{Chandra}}/ACIS-I for 174.5~ks. The
X-ray spectrum is consistent with a single, hyper-luminous, highly
absorbed AGN, and is strongly inconsistent with the presence of a
secondary unobscured AGN. Based on this, we argue that the excess blue
emission in this object is most likely either due to reflection or a
co-eval starburst. We favor the reflection scenario as the unobscured
star-formation rate needed to power the UV/optical emission would be
$\gtrsim 1000~M_{\odot}~\rm yr^{-1}$. Deep polarimetry
observations could confirm the reflection hypothesis.
\end{abstract}


\section{Introduction}\label{sec:intro}

The most luminous infrared (IR) galaxies in the universe are thought
to be massive galaxies during a key stage in their evolution
\citep[see, e.g.,][]{hopkins08}. These galaxies are undergoing a
combination of intense star-formation and intense accretion onto their
central super-massive black hole (SMBH), in both cases heavily
enshrouded in dust. This makes these objects exceedingly luminous in
the IR but typically faint at UV/optical wavelengths. There are
several populations of galaxies that are known to be in such stages,
both locally, such as Ultra-Luminous Infrared Galaxies \citep[ULIRGs;
  e.g,][]{sanders96}, and at high-z, such as Submillimeter Galaxies
\citep[SMGs;][]{blain02,casey14}, Dust-Obscured Galaxies
\citep[DOGs;][]{dey08} and Hot Dust-Obscured Galaxies \citep[Hot
  DOGs;][]{eisenhardt12,wu12}, the latter of which were recently
discovered by NASA's Wide-field Infrared Survey Explorer
\citep[WISE;][]{wright10}.

Hot DOGs have characteristics that clearly separate them from other
populations of luminous galaxies. One of their most distinctive
properties is the hot dust temperatures from which the population
draws its name \cite[see][]{wu12}. Objects such as ULIRGs and SMGs
have IR spectral energy distributions (SEDs) that peak at $\lambda\sim
100~\mu\rm m$ and have typical dust temperatures of up to $\sim 40~\rm
K$ \citep[e.g.,][]{blain03,magnelli12}. This is consistent with their
IR luminosities being powered primarily by star-formation. DOGs can
have somewhat warmer dust temperatures, as they are powered by a
combination of Active Galactic Nuclei (AGN) and star-formation
\citep{melbourne12}. Hot DOGs have much hotter dust temperatures of
$T>60~\rm K$ \citep{wu12}, with shoulders to their SEDs of far-IR
emission very significantly broader than single-temperature modified
black-bodies, and with dust components with temperatures up to $\sim
450~\rm K$ \citep{eisenhardt12,tsai15}. Because of these high
temperatures, their SEDs peak at rest $\lambda\sim 20~\mu\rm m$,
suggesting they are powered by extremely luminous, highly obscured
Active Galactic Nuclei (AGN).

Although the redshift distribution of Hot DOGs is similar to that of
DOGs and SMGs, with most objects found at $1<z<4$
\citep{assef15,eisenhardt15}, their luminosities are significantly
larger. Almost all Hot DOGs seem to have luminosities exceeding
$L_{\rm IR}(8-1000\mu\rm m) = 10^{13}~L_{\odot}$ \citep{tsai15},
placing them in the category of Hyper-Luminous Infrared Galaxies
(HyLIRGs, $L_{\rm IR}>10^{13}~L_{\odot}$). Indeed, a significant
fraction have $L_{\rm IR} > 10^{14}~L_{\odot}$ \citep{wu12}, making
them Extremely Luminous Infrared Galaxies \citep[ELIRGs, $L_{\rm
    IR}>10^{14}~L_{\odot}$;][]{tsai15}. Hot DOGs rank among the
most-luminous galaxies known \citep{tsai15}. In fact, Hot DOG
W2246--0526 is the most luminous galaxy currently known
\citep{tsai15,diaz15}. As expected for such objects, Hot DOGs are a
rare population, with WISE only identifying one candidate every
$\sim30~\rm deg^{2}$ \citep{eisenhardt12,wu12,assef15}, with space
densities comparable to those of similarly luminous QSOs
\citep{assef15}. Hot DOGs also reside in significantly overdense
environments \citep{jones14,assef15} and a large fraction show
extended Ly$\alpha$ emission features on scales of up to $\sim$100 kpc
\citep{bridge13}.

The very red rest-frame optical through mid-IR SEDs of Hot DOGs imply
that the hyper-luminous AGN that powers the IR is under significant
obscuration \citep[$\langle E(B-V)\rangle = 6.4$, up to $E(B-V)\sim
  20$;][]{assef15}, enough that the rest-frame UV-through-NIR SED is
dominated by the underlying stellar emission, providing an
uncompromised view of the host galaxy. Using {\it{Spitzer}} and
ground-based NIR follow-up imaging observations to constrain the
host-galaxy properties, \citet{assef15} studied their stellar masses
and determined upper bounds of $M_{*} \lesssim
10^{12}~M_{\odot}$. They also showed this result is supported by the
density of their environments as measured from {\it{Spitzer}}
follow-up imaging \citep[see][for details]{assef15}. Although Hot DOG
host galaxies may be among the most massive ones at their redshifts,
the AGN is still unexpectedly luminous. \citet{assef15} show that in
order to explain the AGN activity, it is necessary that either the
SMBH is overly massive for its host galaxy or that the AGN is
radiating significantly above the Eddington limit, or possibly a
combination of both. The super-Eddington accretion scenario is also
supported by the study of the most luminous Hot DOGs ($L_{\rm
  IR}>10^{14}~L_{\odot}$) conducted by \citet{tsai15}.

Being able to directly probe the host galaxy of these objects can also
allow us to understand their rest-frame UV/optical
properties. Previous studies of the SEDs of Hot DOGs have shown that
most are well modeled by a regular star-forming galaxy at these
wavelengths \citep[see][as well as
  \S\ref{sec:behds}]{eisenhardt12,assef15}. In this work, however, we
focus on a small fraction of them that show exceedingly blue
UV/optical SEDs. When modeling the SEDs of these Hot DOGs with excess
blue emission, we find with a significant probability that a type 1
AGN is needed to explain the blue rest-frame colors, but with a
bolometric luminosity that is $\sim$1\% of that of the highly obscured
AGN powering the hyper-luminous IR emission. These uncommon sources
may provide important insights into the nature of the Hot DOG
population and their role in the galaxy evolution paradigm. X-ray
observations can provide key tests of explanations for them. As Hot
DOGs are typically faint in soft X-rays, due to the large, possibly
Compton-thick, absorbing H\,{\sc{i}} column density
\citep{stern14,piconcelli15}, deep X-ray observations are required to
carry out such tests.

One of these uncommon sources, WISE J020446.13--050640.8
(\sname\ hereon) is located in the NOAO Deep Wide-Field Survey
\citep[NDWFS]{jannuzi99} Cetus field, within the footprint of the
Large-Area Lyman Alpha survey \citep[LALA;][]{rhoads00}, which has
been observed to a depth of 174.5~ks by the {\it{Chandra X-ray
    Observatory}} with the ACIS-I instrument \citep{wang07}. This
makes \sname\ the Hot DOG with the deepest X-ray observations to
date. We use these X-ray observations in combination with
multi-wavelength data to study \sname. In \S\ref{sec:behds} we discuss
the selection and possible nature of objects with blue excess emission
in general, while in \S\ref{sec:uv_ir_obs} we focus on the UV through
mid-IR SED of \sname. In \S\ref{sec:xrays} we discuss the
{\it{Chandra}} observations, and in \S\ref{sec:discussion} we discuss
the nature of \sname. Throughout this work we assume a flat
$\Lambda$CDM cosmology with $H_0 = 73~\rm km~\rm s^{-1}~\rm Mpc^{-1}$,
$\Omega_M = 0.3$, and $\Omega_{\Lambda} = 0.7$. We refer to all
magnitudes in their standard photometric system, namely AB for $griz$
and Vega for all other bands.

\section{Blue Excess Hot DOGs}\label{sec:behds}

As discussed earlier, the fact that the AGN powering the IR emission
of Hot DOGs is under heavy dust obscuration allows for the possibility
of studying in depth the underlying host galaxy. The rest-frame UV
through NIR SED of a galaxy can provide interesting constraints on the
underlying stellar population, such as its age and mass, as well as
the unobscured star-formation rate \citep[e.g.,][]{kennicutt98} and
the star formation history. \citet{eisenhardt12} showed that the first
Hot DOG studied in depth, WISE J181417.29+341224.9, had an SED
dominated by a strongly star-forming galaxy ($\sim 300~M_{\odot}~\rm
yr^{-1}$) at $\lambda_{\rm rest} \lesssim 1~\mu\rm m$. This is
consistent with the fact that this object has a rest-frame UV spectrum
similar to a Lyman break galaxy, with no discernible emission lines
other than Ly$\alpha$, and several interstellar absorption lines
typical of the Lyman break population
\citep[e.g.,][]{shapley03}. \citet{wu12} showed that although a
fraction of Hot DOGs have similar rest-frame UV spectra, most show
high-ionization narrow emission lines indicative of type 2 AGN and a
small subset even show broad emission lines such as C{\sc iv}. The
expectation is, then, that these objects have a significant range of
rest-frame UV through NIR SEDs.

To study their SEDs we rely here on the subset of sources with
photometry provided by the Sloan Digital Sky Survey
\citep[SDSS;][]{york00}. Note that although Hot DOGs are selected over
the entire extragalactic sky, they are typically faint in the optical,
with $r\gtrsim 23$. This implies that although a significant fraction
of our objects lie within the SDSS footprint, only a fraction are
detected. Specifically we find that 433 out of 934 Hot DOG candidates
fall within the area covered by SDSS DR12 \citep{ahn14}, of which 153
(35\%) are detected with $S/N\ge 3$ in at least one band and 114 are
detected at the same level in at least three passbands. To complement
this, we use the NIR imaging from \citet{assef15} and the deeper
$r$-band observations from the Hot DOG imaging program presented by
\citet{eisenhardt15}. A brief description of the latter is also
provided by \citet{assef15}.

\citet[][also see \citealt{eisenhardt12}]{assef15} recently studied
the rest-frame optical through mid-IR SEDs of a large sample of Hot
DOGs, and showed that the combination of a composite host galaxy SED
template and a single obscured AGN SED template typically does a good
job of modeling the photometry. \citet{assef15} model their SEDs using
the algorithm and templates of \citet{assef10}. The best-fit SED
models consist of a non-negative combination of three empirically
derived galaxy templates, approximately corresponding to E, Sbc and Im
type galaxies, and a single AGN template. They also fit for the
reddening of the AGN component, parametrized by the color excess
$E(B-V)$, considering values from 0 to $10^{1.5}$. The assumed
reddening-law corresponds to an SMC-like extinction for
$\lambda<3300$\AA, and a Galactic extinction curve at longer
wavelengths. Additional details are provided in
\citet{assef10,assef15}. From here on we will refer to this SED model
as the ``1AGN'' model, as a single AGN component is
used. \citet{assef15} modeled the SED of 96 Hot DOGs with
spectroscopic redshifts $z>1$ and for which follow-up imaging had been
obtained by {\it{Spitzer}} \citep{griffith12}, using WISE W3 and W4
bands from the WISE All-sky data release \citep{cutri12}, [3.6] and
[4.5] imaging from the {\it{Spitzer}} follow-up, and $J$, $H$ and $Ks$
imaging from the NIR follow-up program presented by \citet{assef15}
whenever available. The suggested correction to the WISE W3 and W4
band photometry for red sources by \citet[][also see
  \citealt{brown14}]{wright10} was applied before modeling the SEDs.

Here, we recalculate these fits but modify the sample to encompass
only the 36 Hot DOGs with a) $z>1$ \citep[following][]{assef15} and b)
available {\it{ugriz}}
{\tt{modelMag}}\footnote{\url{http://www.sdss.org/dr12/algorithms/magnitudes/\#mag\_model}}
photometry in the SDSS DR12 database with $S/N>3$ in at least one
band. Note that of these, 28 have $S/N\ge 3$ in at least three SDSS
bands. Unlike \citet{assef15}, we also consider here Hot DOGs not
covered by the {\it{Spitzer}} follow-up program, using for them the
lower $S/N$ W1 and W2 fluxes from the AllWISE data release
\citep{cutri13}\footnote{Note that the bias to underestimate the flux
  in the WISE W1 and W2 bands reported by \citet{lake13} for the
  All-sky data release has been fixed in the AllWISE data
  release. Also note that approximately 1\% of Hot DOG candidates are
  not formally detected in the AllWISE catalog, possibly because they
  are confused with fast moving sources. For those, when no
  {\it{Spitzer}} data is available, we use the All-Sky W1 and W2
  fluxes corrected for the bias reported by \citet{lake13}.}
instead. We note that \citet{assef15} decided to not use AllWISE
fluxes due to concerns over the selection function modeling, but these
are not important for our current study. Additionally, we add the
$r$-band photometry from the follow-up program described by
\citet{eisenhardt15} for 25 sources.

\begin{figure}
  \begin{center}
    \plotone{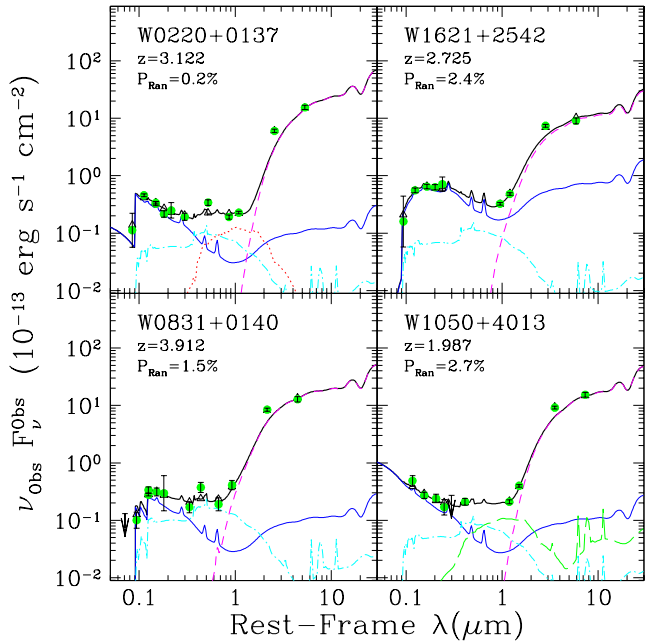}
    \caption{SEDs of the four Hot DOGs with the lowest probability
      $P_{\rm Ran}$ that the improvement in $\chi^2$ gained from
      adding the secondary AGN component is spurious. The green solid
      points show the observed flux densities in a combination of the
      $u^{\prime}$, $g$, $r$, $i$, $z$, $J$, $H$, $Ks$, {\it{Spitzer}}
      and WISE bands (not all bands available for all targets, see
      \S\ref{sec:behds} for details). The best-fit SED model (solid
      black line) consists of a non-negative linear combination of a
      primary luminous, obscured AGN (dashed magenta line), a
      secondary less luminous, unobscured or mildly obscured AGN
      (solid blue line), an old stellar population (dotted red line),
      an intermediate stellar population (dashed green line, not
      needed for all the objects presented here) and a young stellar
      population (cyan dotted-dashed line). Note that not all host
      galaxy templates are used by the algorithm to model the SED of
      each object. The open triangles show the flux density of each
      photometric band predicted by the best-fit SED model.}
    \label{fg:behds_seds}
  \end{center}
\end{figure}

When including the observed optical photometry we find that the
``1AGN'' model no longer provides a good fit for a fraction of Hot
DOGs which present significant excess emission at rest-frame
UV/optical wavelengths compared to what is allowed by the SED
templates of \citet{assef10}. This excess is similar to what would be
expected from an unobscured AGN, although much less luminous than the
one powering the IR emission. To properly identify these sources, we
refit all 36 objects in our sample using the same algorithm described
earlier, but adding an additional AGN component with the same template
for which we fit an independent normalization and reddening value. We
refer to this SED model as the ``2AGN'' model. Note that the ``1AGN''
and ``2AGN'' models have four and six parameters, respectively. An
F-test shows that 8 of the 36 sources (22\%) have good fits and a
probability ($P_{\rm ran}$) below 5\% that the improvement due to the
additional AGN component is spurious. These sources are shown in Table
\ref{tab:behds}\footnote{An additional source (WISE
  J085929.94$+$482302.3) has $P_{\rm ran}=0.034$, but we consider the
  fit to be unreliable. In particular, SDSS reports an $r$-band
  {\tt{modelmag}} flux for this object that differs by over an order
  of magnitude from that measured in a 3\arcsec\ aperture
  ({\tt{fibermag}}), and provides a warning of unreliable
  photometry. The 3\arcsec\ aperture reported by SDSS is consistent
  with the flux we measure in a slightly larger 4\arcsec\ aperture in
  deep $r$-band imaging we have obtained from the WIYN 3m telescope,
  and we can confirm the object is marginally resolved within the
  0.85\arcsec\ PSF of these observations. This suggests the
  {\tt{modelmag}} fluxes reported in the SDSS are unreliable for this
  object and hence we do not consider it further in our
  analysis.}. The four sources with the highest probability of needing
a secondary AGN component are shown, as examples, in Figure
\ref{fg:behds_seds}. The rest-frame UV optical emission is clearly
consistent with that of a type 1 AGN. Because of their significant
excess rest-frame UV/optical emission, we refer to these objects as
Blue Excess Hot DOGs (BHDs).

\begin{deluxetable}{l c c c}

  \tablecaption{Blue Excess Hot DOGs\label{tab:behds}} 

  \tablehead{
    \colhead{WISE ID}&
    \colhead{Redshift}&
    \colhead{W4 (mag)}&
    \colhead{$P_{\rm ran} (10^{-2})$}
  }

  \tabletypesize{\small}
  \tablewidth{0pt}
  \tablecolumns{4}

  \startdata
  WISE J001926.88$-$104633.3 &  1.641 &  7.08 & 3.3\\
  WISE J020446.13$-$050640.8 &  2.100 &  7.06 & 4.5\\
  WISE J022052.12$+$013711.6 &  3.122 &  7.08 & 0.2\\
  WISE J083153.25$+$014010.8 &  3.912 &  7.28 & 1.5\\
  WISE J105045.92$+$401359.1 &  1.987 &  7.08 & 2.7\\
  WISE J131628.53$+$351235.1 &  1.956 &  7.02 & 2.8\\
  WISE J153550.03$+$310054.9 &  1.921 &  6.86 & 2.8\\
  WISE J162101.29$+$254238.3 &  2.725 &  7.66 & 2.4\\
  \enddata

  \tablecomments{The W4 magnitudes presented here correspond to the
    values reported in the All-Sky catalog, uncorrected by the
    \citet{wright10} prescription.}

\end{deluxetable}

While $P_{\rm ran}=5\%$ could be considered a large enough probability
to be wary of the F-test interpretation, we note that the constraints
on the presence of the secondary AGN are much tighter, as the F-test
cannot take into account the fact that the non-negative requirement
for the combination of our templates provides additional
constraints. In order to further assess the need for the secondary AGN
component in the best-fit SED model of these objects, we have created
1,000 realizations of their observed SEDs by re-sampling the
photometry according to its uncertainties. We find that the secondary
AGN component is needed (which we define as the best-fit secondary AGN
component providing at least 2/3 of the flux in one or more of the
SDSS bands) in $98\%$ of the realizations. For comparison, objects
with $P_{\rm ran}>30\%$ use the secondary AGN component in only 66\%
of the realizations.

Determining the fraction of Hot DOGs that show this blue excess is
challenging given the selection effects. As discussed by
\citet{assef15}, the spectroscopic follow-up program has mostly
focused on the brightest objects in W4, namely the 252 Hot DOG
candidates with W4$<$7.2. Of the 36 objects in our sample, 22 fall in
this category and 6 are selected as BHDs. However, there is a strong
bias for BHDs to be detected by SDSS so we instead need to consider
all objects within the SDSS area to assess their true fraction. There
are 51 Hot DOGs with $z>1$ and W4$<$7.2 within the SDSS DR12 area. As
this sample is limited by W4, we posit that objects that are faint at
observed optical wavelengths, and hence reddest in optical--W4, are
unlikely to have blue excesses, and, in turn, be classified as
BHDs. While this is likely a reasonable assumption, this is not
guaranteed since the BHD selection criteria is based primarily on the
UV/optical SED shape, and further observations are needed to properly
assess this. Regardless, assuming that SDSS undetected Hot DOGs are
not BHDs implies a true fraction of $6/51$ (12\%). However there is a
natural bias towards a higher success rate in measuring redshifts for
the optically brighter objects. Considering that 30\% of objects did
not successfully yield a spectroscopic redshift
\citep{assef15,eisenhardt15} and that these are unlikely to be BHDs
for the same reason as discussed above, this implies that the true
fraction is closer to $6/73$, or about 8\% in the W4$<$7.2
sample. Extrapolating this to the entire Hot DOG sample at fainter W4
is not trivial due to the observational biases discussed. We note that
only 2 BHDs are identified among Hot DOGs with W4$>$7.2, suggesting
the true fraction of BHDs might be significantly lower when
considering the entire population.

We suggest three potential scenarios to explain these BHDs.
\begin{itemize}
\item[]{\it{Leaked AGN Light}}: A single AGN is present in the system
  and is responsible for the IR emission by heating the dust that
  surrounds it in almost all directions. However, a small fraction of
  the rest-frame optical/UV emission of this luminous AGN is leaked
  out of the high-obscuration region. One possibility is that a
  fraction of the AGN light is scattered off into our line of
  sight. This effect has been observed in local Seyfert 2 galaxies
  through polarimetry \citep{antonucci85,antonucci93}, and the AGN
  powering the IR emission in Hot DOGs may be luminous enough for its
  scattered component to dominate the rest-frame UV/optical
  fluxes. Alternatively, a small opening between the dust clouds may
  allow us a partial direct line of sight towards the accretion
  disk. The opening would have to be small enough that only $\sim$1\%
  of the accretion disk is directly visible. While in these scenarios
  one would naively expect to see other targets with partially
  obscured UV AGN emission, the SED selection method would not be able
  to identify them as it is only sensitive to those with the largest
  UV excesses.

\item[]{\it{Dual Quasar}}: The secondary AGN emission component may
  come from an additional accreting SMBH in the system that is much
  less luminous than the highly obscured one powering the IR
  emission. Since major mergers are thought to play an important role
  in the evolution of massive galaxies and, in particular, in the
  triggering of intense, obscured AGN and star-formation activity
  \citep[see, e.g.,][]{hopkins08,koss11b}, Hot DOGs could correspond
  to merger stages during which the SMBHs are not yet gravitationally
  bound \citep[e.g.][]{comerford09}. The combination of a dual AGN
  with a luminous, obscured component and a less luminous but
  unobscured one has been previously observed in one candidate and one
  confirmed dual AGN, studied respectively by \citet{barrows12} and
  \citet{fu11}, although the disparity between the component
  luminosities was much smaller than for BHDs. Because dust is likely
  abundant in Hot DOGs even on large scales, we expect the scenario in
  which only one nucleus is unobscured to be more likely if the nuclei
  are separated by scales $\gtrsim 1~\rm kpc$, as this is the typical
  size of the dusty region in ULIRGs \citep{diaz11}.

\item[]{\it{Extreme Star Formation}}: Alternatively, enhanced
  UV/optical emission could be produced by a young, massive
  starburst. A young starburst can have a broad-band power-law shaped
  SED, similar to that of an accretion disk, in the UV/optical regime
  \citep[see, e.g.][]{leitherer99}. The star-formation rate (SFR)
  needed to power such a luminous emission through a young starburst
  would be very substantial. Such an intense starburst would be coeval
  with a hyper-luminous obscured quasar, with important implications
  for galaxy evolution scenarios.

\end{itemize}

Additional observations can help disentangle some of these
scenarios. For example, light leaked from the hyper-luminous AGN
through reflection on the dust grains could be directly probed through
spectropolarimetry, as the characteristic broad emission lines of type
1 AGN should be more apparent in polarized light
\citep{antonucci85,antonucci93}. A young coeval starburst could be
excluded through optical variability, as this would only be expected
for light from an accretion disk. Furthermore, since the typical
amplitude of the variability for a given timescale correlates with the
accretion disk luminosity \citep{vandenberk04,macleod10}, long-term
variability monitoring could allow us to differentiate whether the
excess blue emission light is due to light leaked from the
hyper-luminous AGN or comes from a secondary, lower luminosity active
SMBH. Also, high spatial resolution UV/optical imaging could identify
the two nuclei of a dual AGN, or confirm the presence of a galaxy-wide
massive young starburst. The latter might also be probed by high
resolution ALMA observations of the far-IR continuum and CO emission.

X-ray observations can also help test these scenarios. \citet{stern14}
reported on combined {\it{NuSTAR}} and {\it{XMM-Newton}} observations
of three Hot DOGs without blue excesses, showing that they have
heavily absorbed (possibly Compton-thick) X-ray emission. If BHDs are
due to partial coverage or a secondary unobscured accreting SMBH, we
would expect commensurate unabsorbed X-ray emission coming from the
fainter AGN component. On the other hand, if the excess rest-frame
UV/optical were due to scattered light from the hyper-luminous,
obscured AGN or from extreme star-formation, we would only expect to
see strongly absorbed X-ray emission from the highly obscured AGN
powering the infrared emission as X-ray photons are less scattered by
dust grains or free electrons than UV ones. In principle very deep
X-ray observations can also test the star-formation scenario through a
significant soft X-ray excess \citep[see][and
  \S\ref{ssec:sf}]{mineo14}.

As mentioned in \S\ref{sec:intro}, one of the Hot DOGs that we have
identified as BHDs, namely \sname, is located in the NDWFS Cetus
field, within an area observed by the {\it{Chandra X-ray Observatory}}
ACIS-I instrument. In the following sections we study its
X-ray-through-IR SED to gain more insight into its nature, which may
also help provide insight into the nature of excess blue light from
the other BHDs.

\section{The UV through Mid-IR SED of \sname}\label{sec:uv_ir_obs}

\subsection{Observations}

\sname\ was selected as a Hot DOG candidate through the criteria of
\cite{eisenhardt12}. These candidates (or ``W12-Drops'') are selected
purely by their WISE magnitudes, corresponding to objects with
W1$>$17.4~mag, high $S/N$ detections in either W3 or W4, and with very
red W2--W3 or W2--W4 colors \citep[see][for
  details]{eisenhardt12}. Table \ref{tab:phot} shows the WISE fluxes
for \sname\ from the AllWISE Data Release.

\begin{deluxetable*}{l c c c c c c}

  \tablecaption{Photometry of \sname\label{tab:phot}} 

  \tablehead{
    \colhead{Band}&
    \colhead{Observed $\lambda$} &
    \colhead{Rest $\lambda$} &
    \colhead{Magnitude\tablenotemark{$\dagger$}}&
    \colhead{Flux}&
    \colhead{Telescope/Instrument}&
    \colhead{Ref}\\
    \colhead{}&
    \colhead{($\mu$m)} &
    \colhead{($\mu$m)} &
    \colhead{}& 
    \colhead{($\mu$Jy)}&
    \colhead{}&
    \colhead{}
  }

  \tabletypesize{\small}
  \tablewidth{0pt}
  \tablecolumns{8}

  \startdata
  $g^{\prime}$               &  0.46 &    0.15 &    22.94 (0.05) &     \phn2.4 (0.1)     &           MMT/Megacam & (1)\\
  $r^{\prime}$               &  0.61 &    0.20 &    22.53 (0.06) &     \phn3.5 (0.2)     &           MMT/Megacam & (1)\\
  $i^{\prime}$               &  0.75 &    0.24 &    22.09 (0.09) &     \phn5.2 (0.4)     &           MMT/Megacam & (1)\\
  $z^{\prime}$               &  0.89 &    0.29 &    21.69 (0.06) &     \phn7.4 (0.4)     &           MMT/Megacam & (1)\\
  $u$                        &  0.35 &    0.13 &    23.00 (0.60) &     \phn2.2 (1.3)     &                  SDSS & (2)\\
  $g$                        &  0.46 &    0.15 &    22.66 (0.17) &     \phn3.1 (0.5)     &                  SDSS & (2)\\
  $r$                        &  0.61 &    0.20 &    22.49 (0.23) &     \phn3.6 (0.8)     &                  SDSS & (2)\\
  $i$                        &  0.75 &    0.24 &    21.80 (0.18) &     \phn6.9 (1.2)     &                  SDSS & (2)\\
  $z$                        &  0.89 &    0.29 &    22.03 (0.67) &     \phn4.3 (4.3)     &                  SDSS & (2)\\
  $r$                        &  0.62 &    0.20 &    22.36 (0.17) &     \phn4.1 (0.6)     &              SOAR/SOI & (3)\\
  $J$                        &   1.2 &    0.40 &    20.77 (0.22) &     \phn8.0 (1.6)     &            WIYN/WHIRC & (4)\\
  $[3.6]$                    &   3.6 &    1.15 &    17.18 (0.05) &         37.2 (1.9)    &   {\it{Spitzer}}/IRAC & (5)\\
  $[4.5]$                    &   4.5 &    1.45 &    16.34 (0.03) &         52.2 (2.6)    &   {\it{Spitzer}}/IRAC & (5)\\
  W1                         &   3.4 &    1.06 &    17.34 (0.12) &         35.4 (3.8)    &                  WISE & (6)\\
  W2                         &   4.6 &    1.47 &    16.10 (0.16) &         61.8 (9.0)    &                  WISE & (6)\\
  W3\tablenotemark{$\ddagger$} &    12 &    3.41 &    10.25 (0.06) &       2714 (140)      &                  WISE & (6)\\
  W4\tablenotemark{$\ddagger$} &    22 &    7.07 & \phn7.07 (0.09) &      11414 (946)\phn  &                  WISE & (6)\\
  P70                        &    70 &   22.58 &    \nodata      &      71000 (3000)     & {\it{Herschel}}/PACS  & (7)\\
  P160                       &   160 &   54.84 &    \nodata      &     113000 (5000)\phn & {\it{Herschel}}/PACS  & (7)\\
  S250                       &   250 &   80.65 &    \nodata      &  \phn46000 (11000)    & {\it{Herschel}}/SPIRE & (7)\\
  S350                       &   350 &  112.90 &    \nodata      &  \phn27000 (11000)    & {\it{Herschel}}/SPIRE & (7)\\
  S500                       &   500 &  161.29 &    \nodata      &  \phn17000 (15000)    & {\it{Herschel}}/SPIRE & (7)\\
  \enddata

  \tablecomments{References: (1) \citet{finkelstein07}; (2)
    \citet{ahn14}; (3) \citet{eisenhardt15} (4) \citet{assef15}; (5)
    \citet{griffith12}; (6) \citet{wright10}; (7) \citet{tsai15}}
 
  \tablenotetext{$\dagger$}{Magnitudes are presented in their standard
    photometric system, namely AB for
    $g^{\prime}r^{\prime}i^{\prime}z^{\prime}$ and $ugriz$, and Vega
    for the rest. For the SDSS bands we present $asinh$ magnitudes,
    while we present Pogson magnitudes for all others.}

  \tablenotetext{$\ddagger$}{As discussed in \S\ref{sec:behds}, we have
    corrected the W3 and W4 fluxes according to the prescription
    suggested by \citet{wright10} for objects with red WISE colors.}

\end{deluxetable*}

By selection, Hot DOGs have low $S/N$ W1 and W2 fluxes. These
wavelengths, however, provide significant information about the host,
and hence are crucial for modeling the SEDs of these sources. To this
end, we obtained {\it{Spitzer}} imaging of \sname\ in the [3.6] and
[4.5] bands on UT 2011 February 28, as part of a comprehensive survey
of the Hot DOG population \citep{griffith12}. We also obtained
$J$-band imaging of \sname\ using the WHIRC camera on the WIYN
telescope on the night of UT 2012 January 1 as part of the NIR
follow-up campaign \citep{assef15}. Although optical photometry is
available for this source from SDSS DR12 and the follow-up program
mentioned in the previous section, we use instead deeper
$g^{\prime}r^{\prime}i^{\prime}z^{\prime}$ imaging obtained by
\citet{finkelstein07} using the Multiple Mirror Telescope/Megacam in
2005 November and 2006 January. The fluxes of all these bands are
listed in Table \ref{tab:phot}. The profile of this object is
dominated by a central point source but shows an extended component
that accounts for up to 30\% of the integrated flux. We note that all
three measurements of $r$-band photometry are consistent within the
error-bars, so we cannot attempt to use optical variability to test
the proposed scenarios (see \S\ref{sec:behds}). Table \ref{tab:phot}
also shows the fluxes of \sname\ in the {\it{Herschel}}/PACS 70$\mu$m
and 170$\mu$m bands and in the {\it{Herschel}}/SPIRE 250, 350 and
500$\mu$m bands, obtained as part of a Hot DOG follow-up program
\citep[PID: {\tt{OT2\_peisenha\_2}}, PI: Eisenhardt; see][for
  details]{tsai15}.

\begin{figure}
  \begin{center}
    \plotone{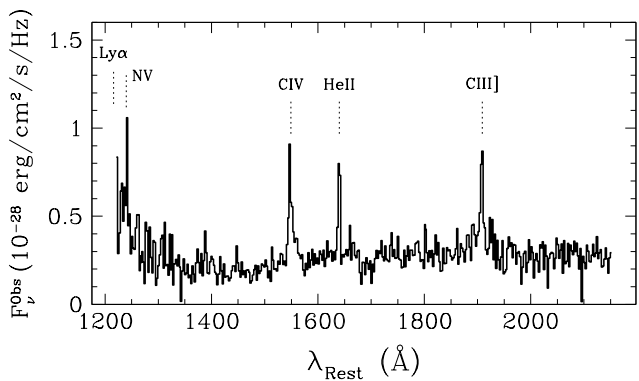}
    \caption{Gemini South/GMOS-S optical spectrum of \sname, binned in
      wavelength by a factor of 8 for clarity. The dotted vertical
      lines show the wavelength of the Ly$\alpha$, N{\sc v}, C{\sc
        iv}, He{\sc ii} and C{\sc iii}] emission lines. The detected
    emission lines yield a redshift of $z=2.100$.}
    \label{fg:gmos_spec}
  \end{center}
\end{figure}

An optical spectrum of \sname\ was obtained with the GMOS-S
spectrograph on the Gemini South telescope on UT 2011 November 27,
using the 1.5\arcsec\ $\times$ 108\arcsec\ longslit and the
B600\_G5323 disperser with a 2$\times$600s exposure time as part of an
optical spectroscopic follow-up campaign of Hot DOGs
\citep{eisenhardt15}. Figure \ref{fg:gmos_spec} shows the reduced
spectrum of \sname, which displays the \CIV, \HeII\ and
\CIII\ emission lines at a redshift of $z=2.100\pm 0.002$. Each of the
three emission lines is statistically well modeled by a single
Gaussian component given the $S/N$ of our observations. For \civ,
\heii\ and \ciii\ we find that the best-fit Gaussian components have
rest-frame FWHM of $1630\pm 220$, $950\pm 200$ and $550\pm 100~\rm
km~\rm s^{-1}$ respectively. Note that the \heii\ emission line is
comparable in strength to \civ\ and \ciii, which is atypical of
broad-lined quasars. For example, in the SDSS composite quasar
spectrum of \citet{vandenberk01}, \heii\ has 2--3\% the strength of
\civ\ and \ciii\, though the comparison is unfair since \heii\ is
narrow and the carbon lines are dominated by broad emission in
unobscured sources. For high luminosity narrow-line AGN such as radio
galaxies, \heii\ is often comparable in strength to the carbon lines,
as seen in the composite radio galaxy spectra of \citet{mccarthy93}
and \citet{stern99}. Furthermore, the FWHM we find for \civ\ and
\heii\ are consistent with those measured in the composite spectrum of
\citet[][respectively 1540 and $1150~\rm km~\rm s^{-1}$]{stern99},
although \ciii\ is significantly narrower than the $1260~\rm km~\rm
s^{-1}$ found by \citet{stern99}. As another example of the
similarities between Hot DOGs and radio galaxies, comparing
{\it{Spitzer}} imaging of Hot DOGs to those of radio galaxies as
reported by \citet{wylezalek13}, \citet{assef15} showed that Hot DOGs
reside in similarly overdense environments, suggestive of
high-redshift proto-clusters in the process of formation.

Combining the redshift with the {\it{Spitzer}}, WISE and
{\it{Herschel}} observations we estimate an infrared luminosity of
$L_{\rm IR} = 4.4\times 10^{13}~L_{\odot}$ using the approach of
\citet{tsai15}, which puts \sname\ well into the HyLIRG
category. \sname\ is not detected by the VLA FIRST survey at 1.4~GHz,
with a catalog detection limit of 1~mJy/beam. Assuming the object is a
point source, this translates into a specific luminosity limit of
$L_{\nu}(4.34~\rm GHz) < 9.7\times 10^{31}~\rm erg~\rm s^{-1} ~\rm
Hz^{-1}$ and a limiting flux density ratio between rest-frame 4.34~GHz
and rest-frame 4400\AA\ of $f_{\nu}(4.34~\rm GHz)/f_{\nu}(4400\rm \AA)
< 1000$, neither of which is stringent enough to classify \sname\ as
radio-loud nor confirm it as radio-quiet \citep{stern00}.

\subsection{SED Modeling}\label{ssec:sed_model}

\begin{figure}
  \begin{center}
    \plotone{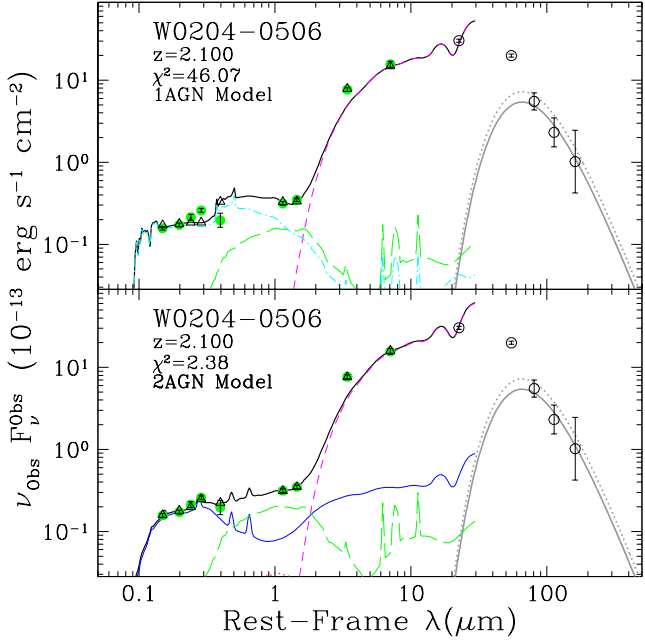}
    \caption{UV through mid-IR SED of \sname. The solid green circles,
      open triangles, solid black line and color lines have the same
      meaning as in Fig. \ref{fg:behds_seds}. The top and bottom
      panels show the best-fit ``1AGN'' and ``2AGN'' models to the SED
      (see \S\ref{sec:behds} for details). The ``2AGN'' model clearly
      provides a better description of the observed flux densities.
      The open circles show the flux densities of \sname\ in the
      {\it{Herschel}} band; these were not used in the SED
      modeling. The solid gray line shows the best-fit modified
      blackbody (assuming $\beta=1.5$ and $T=40~\rm K$; see
      \S\ref{ssec:sf} for details) to the three longer wavelength
      (SPIRE) bands. The dotted gray line shows the same but with a
      luminosity at the 90\% confidence limit above that of the
      best-fit.}
    \label{fg:sed}
  \end{center}
\end{figure}

The top panel of Figure \ref{fg:sed} shows the best-fit ``1AGN'' model
to the UV-through-mid-IR photometry of \sname. The fit to the optical
photometry is poor, as reflected by the large value of $\chi^2=46.07$
with only four degrees of freedom for the fit ($\chi^2_{\nu} =
11.5$). The bottom panel of this Figure shows the best-fit ``2AGN''
model, which provides a significantly better $\chi^2$ value of 2.38
($\chi^2_{\nu} = 1.2$). Comparing both models through an F-test yields
$P_{\rm Ran} = 5\times 10^{-2}$, implying that the addition of a
secondary unobscured AGN component is justified (see discussion in
\S\ref{sec:intro}). The best-fit ``2AGN'' model consists of a highly
obscured luminous AGN that dominates the infrared luminosity with
$E(B-V)=9.7\pm 1.2$ and a 6$\mu$m luminosity of $L_{6\mu\rm m} =
1.9\pm 0.2\times 10^{13}~L_{\odot}$. The rest-frame UV/optical is, on
the other hand, dominated by a lightly reddened AGN component with
$E(B-V)=0.13\pm 0.02$ and a significantly lower 6$\mu$m luminosity of
$L_{6\mu\rm m} = 2.7\pm 0.5 \times 10^{11}~L_{\odot}$, of order 1\% of
the luminosity of the highly obscured AGN.

SED modeling can also allow us to constrain the stellar mass of this
object. Unfortunately, we do not have enough information to constrain
the $M/L$ ratio, so instead we use the approach of \citet{assef15} and
estimate an upper bound. The maximal stellar mass estimate analysis
for Hot DOGs done by \citet{assef15} yields $M^{\rm Max}_{*} = 7\times
10^{11}~M_{\odot}$ for this object, assuming the ``1AGN'' model. The
``2AGN'' best-fit will yield a lower upper-bound on the stellar mass,
as the rest-frame $K$-band luminosity is less dominated by the stellar
emission, so we use the estimate of the ``1AGN'' model as it is the
most conservative upper-bound.

While the ``2AGN'' model fits the data very well, we cannot
immediately interpret this as proving the existence of the secondary
unobscured AGN component in the SED of \sname\ expected for the dual
AGN and reflection scenarios. Instead, the blue excess emission may be
described by an extremely luminous and young starburst, as described
in \S\ref{sec:behds}, which falls beyond the parameter space covered
by the empirical host galaxy templates we considered. In the next
section we study the deep X-ray observations available for this object
to gain further insight into the nature of the unusual SED of \sname.

\section{X-ray Observations of \sname}\label{sec:xrays}

\sname\ was serendipitously observed by the {\it{Chandra X-ray
    Observatory}} as part of the LALA Cetus field observations (PID:
04700805, PI:Malhotra) presented by \citet{wang07}. The field was
observed during Cycle 4 with the ACIS-I \citep{garmire03} instrument
for 160~ks starting on UT 2003 June 13 and then again for 15~ks more
starting on UT 2003 June 15. For simplicity we only use the 160~ks
observation. The {\it{Chandra}} observation was taken in the Timed
Event mode, and we extracted spectra from the ACIS-I detector using
the standard pipeline in CIAO
v4.6\footnote{\url{http://cxc.harvard.edu/ciao/}}. The source spectrum
was obtained from a circular region of radius $\sim$2\arcsec, while
the background was extracted from a larger nearby circular region that
was free from any other contaminating sources. The source is detected
with 94 counts and is visually consistent with a point source. Owing
to the low count statistics, the source spectrum was only lightly
grouped with a minimum of 1 count per bin. We therefore carry out the
parameter estimation by minimizing the Cash statistic \citep{cash79},
modified through the W-statistic provided by
XSPEC\footnote{\url{https://heasarc.gsfc.nasa.gov/xanadu/xspec/manual/XSappendixStatistics.html}}
to account for the subtracted background.

\begin{figure}
  \begin{center}
    \plotone{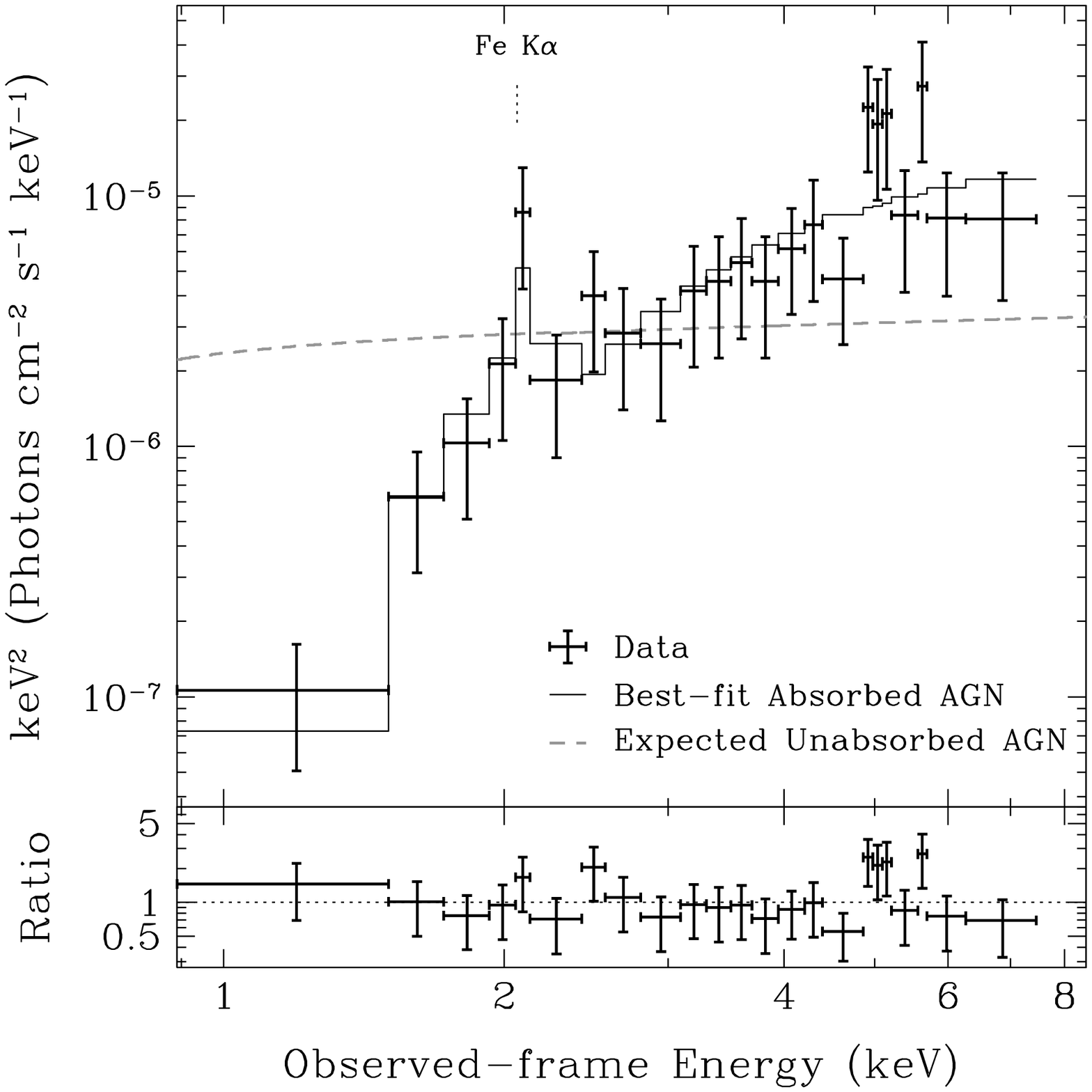}
    \caption{({\it{Top}}) The black crosses show the binned, unfolded
      {\it{Chandra}}/ACIS-I spectrum of \sname\ from the LALA field
      observations of \citet{wang07}. The data have been rebinned here
      for visual purposes only. The solid black histogram shows the
      best-fit absorbed AGN using the models of
      \citet{brightman11}. The dashed gray line shows the spectrum
      expected for the lightly obscured AGN component suggested by the
      excess blue emission of \sname. ({\it{Bottom}}) Flux ratio
      between the observations and the best-fit absorbed AGN. }
    \label{fg:xray_spec}
  \end{center}
\end{figure}

Figure \ref{fg:xray_spec} shows the unfolded spectrum from the
{\it{Chandra}} observations. The emission is clearly hard, implying it
is powered by a highly absorbed AGN, as expected from the
multi-wavelength SED presented in the previous section. While the
Figure may show a tentative Fe K$\alpha$ line, the counts are too low
to determine whether the line is real or simply a statistical
fluctuation. The Figure also shows the best-fit absorbed AGN obtained
using the models of \citet{brightman11}.  These models predict the
X-ray spectrum as observed through an optically thick medium with a
toroidal geometry, as posited by the AGN unified scheme. The models
employ Monte-Carlo techniques to simulate the transfer of X-ray
photons through the optically-thick neutral medium, self-consistently
including the effects of photoelectric absorption, Compton scattering
and fluorescence from Fe K, amongst other elements. Treating these
effects self-consistently rather than separately has the advantage of
reducing the number of free parameters and of gaining constraints on
the spectral parameters. It is therefore particularly useful for low
count spectra such as the one we are fitting here.

The best-fit model yields a photon-index of
$\Gamma=1.6^{+0.8}_{-0.0}$, a neutral hydrogen column density of
$N_{\rm H}=6.3^{+8.1}_{-2.1} \times 10^{23}~\rm cm^{-2}$ and an
absorption-corrected 2--10~keV luminosity of $\log L_{2-10\rm keV} /
\rm erg~\rm s^{-1} = 44.9^{+0.86}_{-0.14}$.  The best-fit model has a
Cash statistic of $C=66.08$ for 77 degrees of freedom. The
uncertainties quoted correspond to the 90\% confidence interval. Note
that the photon-index is poorly constrained by the data, so we require
$\Gamma\ge 1.6$ since lower values of the photon-index are only
observed in SMBHs accreting at low Eddington rates
\citep[e.g.,][]{shemmer06,shemmer08,risaliti09,brightman13}.

\begin{figure}
  \begin{center}
    \plotone{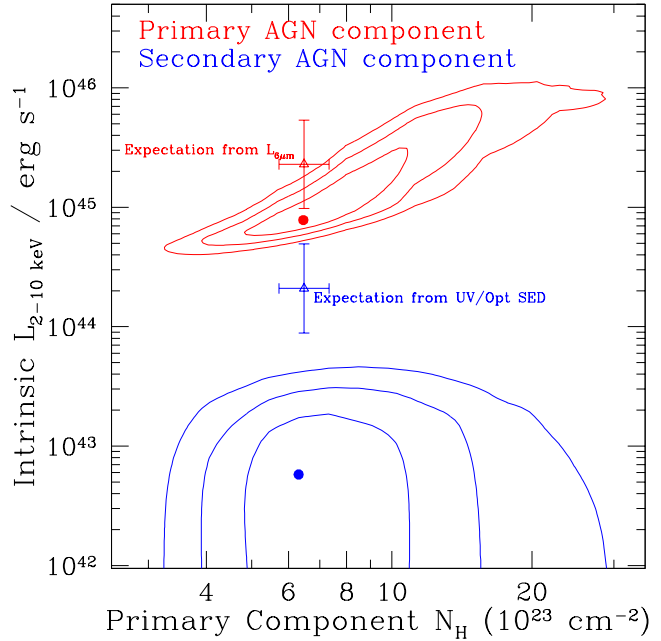}
    \caption{Confidence intervals of the joint fit of the two AGN
      components to the {\it{Chandra}}/ACIS-I spectrum as described in
      the text. The red lines show the 1, 2 and 3$\sigma$ confidence
      contours of the best-fit $L_{2-10~\rm keV}$ and $N_{\rm H}$
      parameters for the primary (i.e., highly luminous, highly
      absorbed) AGN component. The photon-index, $\Gamma$ was also fit
      for this component. The blue lines show the 1, 2 and 3$\sigma$
      confidence contours of the best-fit $L_{2-10~\rm keV}$ of the
      secondary (i.e., less luminous, lightly absorbed) AGN component
      as a function of the $N_{\rm H}$ of the primary AGN
      component. The $N_{\rm H}$ and $\Gamma$ of the secondary
      component were fixed to $8.9\times 10^{21}~\rm cm^{-2}$ and to
      1.9 respectively, as described in the text. The red and blue
      open triangles show expected values obtained from the best-fit
      ``2AGN'' model as described in the text.}
    \label{fg:lx_nh}
  \end{center}
\end{figure}

The properties derived for the luminous, highly absorbed AGN emission
dominating the X-rays are in good agreement with those derived for the
luminous, highly obscured AGN component that dominates the IR
SED. Assuming the median dust-to-gas ratio observed by
\citet{maiolino01} in AGN, namely $E(B-V)/N_{\rm H}=1.5\times
10^{-23}~\rm cm^{2}~\rm mag$, the best-fit accretion disk obscuration
found through the SED modeling for the highly obscured AGN corresponds
to a neutral hydrogen column density of $N_{\rm H} = 6.5\pm 0.8 \times
10^{23}~\rm cm^{-2}$, a value which is consistent with that measured
from the X-ray spectrum. Similarly, we can compare the estimated
luminosities, as the intrinsic $L_{2-10\rm keV}$ luminosity of an AGN
correlates well, although with considerable scatter, with its
monochromatic luminosity at 6$\mu$m, $L_{6\mu\rm m}$
\citep[e.g.,][]{fiore09,gandhi09,bauer10,mateos15,stern15}. These
relations are typically linear \citep[with the exception of][who used
  a broken power-law]{fiore09} and have been derived for limited
energy regimes, but \citet{stern15} has recently shown that a
quadratic relation does a better job at describing the entire AGN
luminosity range. In particular, this relation has been derived
considering objects with luminosities as high as those of Hot DOGs,
making it the most appropriate one to use in this context. From the
relation presented by \citet{stern15}, adding its scatter of 0.37~dex
to the uncertainty budget, we get that the best-fit $L_{6\mu\rm m}$
for the highly obscured, luminous AGN component corresponds to an
intrinsic X-ray luminosity of $\log L_{2-10\rm keV} / \rm erg~\rm
s^{-1} = 45.36\pm 0.37$, which is consistent with the estimate from
the X-ray spectrum. Figure \ref{fg:lx_nh} shows the confidence
contours for $\log L_{2-10\rm keV}$ and $N_{\rm H}$ obtained from the
X-ray spectrum, as well as the estimates based on the multi-wavelength
SED described above. The latter are consistent with the X-ray spectrum
estimates, although the agreement is better once the scatter of the
$L_{2-10\rm keV} - L_{6\mu\rm m}$ is taken into account. This
consistency implies that the same AGN component is responsible for
both the IR SED and the X-ray emission, as expected.

In \S\ref{sec:behds} we suggested three mechanisms to explain the blue
excess component that dominates the rest-frame UV/optical SED of
BHDs. For the presence of a secondary active SMBH with little or no
obscuration, we would expect to see the corresponding
unabsorbed/lightly absorbed X-ray emission. However, the X-ray
spectrum in Figure \ref{fg:xray_spec} is strongly inconsistent with
the presence of an additional AGN component with little or no
absorption. If we use the $L_{6\mu\rm m}-L_{2-10\rm keV}$ relation of
\citet{stern15}, we find that the corresponding X-ray luminosity for
the secondary AGN that best-fits the rest-frame UV/optical SED would
be $\log L_{2-10\rm keV}/\rm erg~\rm s^{-1}= 44.32\pm 0.37$. Assuming
the median gas-to-dust ratio of \citet{maiolino01}, the best-fit
obscuration of $E(B-V)=0.13$ corresponds to a low absorption of
$N_{\rm H} = 8.9\times 10^{21}~\rm cm^{-2}$; such a component would
have significant power at soft energies. Figure \ref{fg:xray_spec}
shows the expected X-ray signal from a power-law spectrum with the
above luminosity and a photon-index $\Gamma=1.9$ under this light
amount of absorption in comparison to the observed X-ray spectrum,
illustrating that the observations are highly inconsistent. If we
allow for the luminosity of this component to be fit to the spectrum
(keeping $N_{\rm H}$ and $\Gamma$ fixed) simultaneously with the
parameters for the heavily absorbed power-law described earlier, we
find that with 90\% confidence $\log L_{2-10\rm keV} / \rm erg~\rm
s^{-1} < 43.3$, over an order magnitude below the expectation from the
luminosity of the component that dominates the rest-frame UV/optical
SED. Figure \ref{fg:lx_nh} shows the confidence intervals for the
luminosity of this component as well as the expectation from the
UV/optical SED signal. The confidence intervals for the primary
component are qualitatively unaffected by the additional unobscured
component and hence are not shown. Fixing the X-ray luminosity of the
component to the expected value of $\log L_{2-10\rm keV}/\rm erg~\rm
s^{-1}= 44.32$ yields a Cash statistic of $C=194.46$. This implies a
$\Delta\rm C=128.38$ above the best-fit, which means we can rule this
scenario out with $>99.9\%$ confidence.

\section{The Origin of the Blue Excess Emission in \sname}\label{sec:discussion}

As stated in \S\ref{sec:behds}, there are three scenarios that can
naturally explain the SED of \sname. Below we discuss each of them in
light of the observations described in the previous sections.

\subsection{Leaked AGN Light}\label{ssec:leaked_light}

As discussed in \S\ref{sec:behds}, a source for the blue, AGN-like
rest-frame UV/optical SED could be that the UV emission from the
hyper-luminous, highly obscured AGN that powers the infrared emission
of \sname\ is leaking out from the inner regions of the galaxy. This
could happen, in principle, from scattering of the central engine
light by free electrons, or by reflection on dust grains. The fraction
of the emitted flux from the obscured central engine that is leaking
into our line of sight in the UV/optical would be of order 1\%, as
that is the relative luminosity found by the SED modeling between the
intrinsic luminosity estimated for the AGN powering the rest-frame
UV/optical to that needed to power the IR emission. The cross section
for scattering by either free electrons or by dust grains, however, is
significantly smaller in the energy range of our ACIS-I spectrum than
in the UV \citep{draine03a,draine03b}, so the lack of a luminous soft
X-ray component found in \S\ref{sec:xrays} is consistent with this
scenario. This makes the reflection the most likely scenario we
review, although we note that this simple estimate neglects a
dependence of reflection on wavelength across the SED. For reflection
off dust grains, such an effect depends on the specific properties of
the dust grains and on the geometry of the dust with respect to SMBH,
but full modeling of it falls well beyond the scope of this paper.

An alternative method for the emission to escape from the
hyper-luminous, highly obscured AGN would be to simply have a
partially unobscured line-of-sight towards the accretion disk, such
that 1\% of the emission is reaching us directly. The fact that we see
emission consistent with a lightly obscured accretion disk up to the
Lyman-break already makes this unlikely. The partial coverage would
have to allow a direct line of sight to 1\% of the accretion disk at
all wavelengths, despite the rest-frame UV coming primarily from
regions closer in to the BH than the longer optical wavelengths
\citep{shakura73,kochanek04,anguita08}, and from a physically distinct
region than the X-rays \citep{vaiana78,haardt93}. Figure
\ref{fg:lx_nh} shows that the X-ray data cannot rule out a secondary
unobscured AGN component with 1\% the X-ray luminosity of the obscured
component, as would be expected for this partial obscuration
scenario. Furthermore, it is not necessary for the gas and dust
distribution to trace each other perfectly \citep[see,
  e.g.,][]{merloni14}. Nonetheless, the complex dust geometry that
would be necessary to obtain the observed SED in this scenario makes
it unlikely.

\subsection{Dual Quasar}\label{ssec:dual_agn}

Alternatively, the unobscured AGN emission in the rest-frame
UV/optical SED could be from an independent accreting SMBH, making
\sname\ a dual quasar system. Dual AGN are rare objects, but there are
several confirmed cases in the literature
\citep[e.g.][]{komossa03,hudson06,bianchi08,
  koss11,fu11,comerford11,liu13}. Furthermore, this scenario might
      {\it{a priori}} be plausible, as hyper-luminous AGN activity
      phases such as those in Hot DOGs might be triggered by major
      galaxy mergers \citep[e.g.,][]{hopkins08}. This scenario is,
      however, inconsistent with the lack of a lightly absorbed AGN
      component at the expected luminosity in the X-ray spectrum, as
      discussed in \S\ref{sec:xrays}. This scenario could fit the
      observations if the secondary, lightly obscured AGN was
      intrinsically X-ray weak \citep{luo13,luo14,teng14}, but all
      such objects identified to date are broad absorption line (BAL)
      QSOs, and we do not see any trace of BAL features in the optical
      spectra (see Fig. \ref{fg:gmos_spec}). Recently \citet{teng15}
      has pointed out that AGN in ULIRGs may be potentially X-ray weak
      too, although it is hard to tell how much of a role obscuration
      plays in these objects. Hence, while we cannot completely rule
      out the dual quasar scenario, we consider it unlikely given the
      ACIS-I observations presented here.

\subsection{Extreme Star-Formation}\label{ssec:sf}

It is possible that the UV/optical excess observed in BHDs is not due
to accretion onto an SMBH, but to intense, unobscured
star-formation. Over the first $\sim$100~Myr, a pure starburst has a
UV/optical SED rising strongly towards the blue, up to the Lyman
break. This is similar to the SED of unobscured QSOs. Hence, it is
possible to model the SED of \sname\ as a mildly obscured young
starburst instead of as a mildly obscured AGN in
\S\ref{sec:uv_ir_obs}. Detailed modeling is not feasible with the low
number of photometric bands available for \sname\ in the rest-frame
UV/optical, so we concentrate here in determining the lowest possible
amount of star-formation that would be needed to power its UV/optical
SED. We use SED models generated with the Starburst99 v7.0.0 code
\citep{leitherer99,leitherer10,leitherer14,vazquez05} in combination
with the {\it{EzGal}} package of \citet{mancone12}, and take the
following approach.

For \sname, $J$-band corresponds to rest-frame 4000\AA, so its blue
$z^{\prime}-J$ color precludes the presence of a strong Balmer-break,
implying a very young age for the starburst. We assume models with a
constant SFR generated by Starburst99, although our results are
qualitatively similar if we instead use a simple stellar
population. We assume the latest Geneva models available for this
version of Starburst99 \citep[see][for details]{leitherer14}. Since
the $g^{\prime}-r^{\prime}$ color is significantly redder than would
be expected for a young starburst, we also allow for obscuration to be
fit to this component, assuming the same reddening-law as in
\S\ref{sec:uv_ir_obs}. Assuming a reddening law with a strong
2175\AA\ feature, as is observed in the Milky Way
\citep[e.g.,][]{cardelli89}, results in a poorer description of the
SED. The large amount of dust present in the system is indicative of a
considerable mean metallicity, so we assume a solar metallicity for
our models, although we discuss the effects of this choice later in
this section. We assume a Salpeter initial mass function
\citep[IMF;][]{salpeter55} with mass ranges between $0.08~M_{\odot}$
and $120~M_{\odot}$. For the rest of the input parameters of the
Starburst99 code we assume the standard recommended options.

\begin{figure}
  \begin{center}
    \plotone{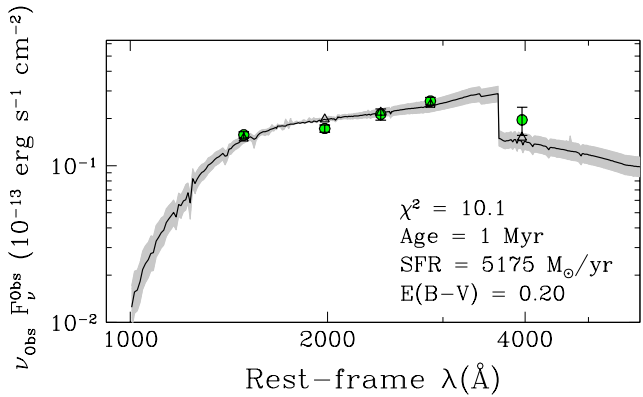}
    \caption{The solid black line shows the best-fit SED model from
      Starburst99 to the $g^{\prime}$, $r^{\prime}$, $i^{\prime}$,
      $z^{\prime}$ and $J$ band flux densities (solid green
      circles). The expected flux densities from the best-fit model
      are shown by the open triangles, and its physical parameters are
      given in the legend. The gray region shows all SED shapes within
      the 90\% confidence interval.}
    \label{fg:sf_fit}
  \end{center}
\end{figure}

\begin{figure}
  \begin{center}
    \plotone{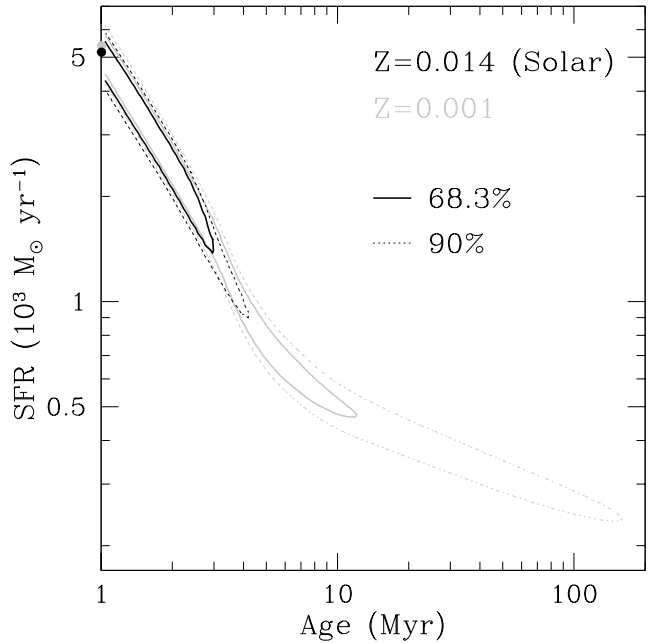}
    \caption{The contours show a $\chi^2$ map of the best-fit SED
      models of Starburst99 to the $g^{\prime}$, $r^{\prime}$,
      $i^{\prime}$, $z^{\prime}$ and $J$ band flux densities for a
      fixed age and SFR of the starburst (see \S\ref{ssec:sf} for
      details of the SED modeling). The contours for the model
      assuming a solar metallicity ($Z=0.014$) are shown in black
      while the contours for the model assuming $Z=0.001$ are shown in
      gray. The solid (dotted) contour shows the 68.3\% (90\%)
      confidence region. The solid dots show the best-fit models (see
      Fig. \ref{fg:sf_fit}).}
    \label{fg:ages_SFR_contour}
  \end{center}
\end{figure}

Figure \ref{fg:sf_fit} shows the best-fit SED to the observed
UV/optical SED. When fitting the SED, we impose a minimum age of
$1~\rm Myr$ and find the best-fit model has $\chi^2=10.1$, with an age
of $1~\rm Myr$, $E(B-V) = 0.20$ and $\rm SFR = 5200~M_{\odot}~\rm
yr^{-1}$. Note that the fit is much poorer than for the AGN model,
particularly as we only consider five data points. As we fit for the
amplitude, reddening and age, this implies $\chi^2_{\nu} = 10.1/2 =
5.05$. If we lift the minimum age requirement we find the best-fit SED
requires an enormous SFR of approximately $50,000~M_{\odot}~\rm
yr^{-1}$ but with an almost indistinguishable $\chi^2=9.9$. This
implies there is a strong degeneracy between age and SFR in our
models, which is made clearer by the contours shown in Figure
\ref{fg:ages_SFR_contour}. Considering this degeneracy, we can
determine that, with 90\% confidence, $\rm SFR\gtrsim
1000~M_{\odot}~\rm yr^{-1}$. While $\rm SFR>1000~M_{\odot}~\rm
yr^{-1}$ is routinely observed in SMGs and ULIRGS
\citep[e.g.,][]{barger14}, it is always heavily dust obscured and
hence only detected in the far-IR. \citet{barger14} shows, based on
the measurements of \citet{vanderburg10}, that UV measured SFRs in
Lyman break galaxies at $z>3$ cut off at $\sim 300~M_{\odot}~\rm
yr^{-1}$, suggesting that an unobscured $\rm SFR_{\rm Min} \sim
1000~M_{\odot}~\rm yr^{-1}$ is very unlikely.

A somewhat better fit to the data can be obtained by assuming a lower
metallicity. If we consider the lowest possible metallicity provided
for the Geneva models by Starburst99, namely $Z=0.001$, we find that
the best-fit SED has $\chi^2=7.8$ ($\chi^2_{\nu} = 3.9$) and $\rm SFR
= 5400~M_{\odot}~\rm yr^{-1}$. Yet, as shown in Figure
\ref{fg:ages_SFR_contour}, this significantly relaxes the requirement
on the minimum SFR, implying only $\rm SFR\gtrsim 250~M_{\odot}~\rm
yr^{-1}$ with 90\% confidence. This would be consistent with the upper
envelope of what has been found for Lyman break galaxies. However this
model would require that the gas feeding the starburst has far lower
metallicity than that feeding and surrounding the nucleus, so we do
not consider it any further.

Note that we have not included the WISE W1 band flux to constrain
these fits despite the fact that W1 would be dominated by the host
galaxy emission if the blue-excess is due to star-formation (see
Fig. \ref{fg:sed}). Adding the W1 flux as a constraint results in a
much poorer fit by our constant SFR model ($\chi^{2} = 28$) that
severely underestimates the observed W1 flux. This is not surprising,
as it simply highlights the need for an older stellar population in
the system. However, it is worth noting that adding an older stellar
component would only make the best-fit SFR larger and the starburst
younger, as it would imply an even stronger inverse Balmer-break for
the starburst. Nevertheless, we expect the changes to be negligible
based on the W1 amplitude and the large $J$-band flux uncertainty.

An independent constraint on the SFR can be obtained from
{\it{Herschel}}/SPIRE observations of \sname, as the cold dust
emission from star-forming regions is a good tracer of their activity
\citep[e.g.][]{kennicutt98}. As discussed by previous studies
\citep{eisenhardt12,wu12,wu14,jones14,tsai15}, the far-IR SEDs of Hot
DOGs are dominated by the hot dust emission powered by the AGN, and
show no evidence for a significant cold component associated with
star-formation. Assuming that the {\it{Herschel}}/SPIRE fluxes of
\sname\ are powered solely by star formation, we can thus place a
conservative limit on the SFR. We fit the 250, 350 and 500$\mu$m
fluxes with a modified blackbody and then estimate the $\rm SFR$ limit
using the $L_{\rm FIR}-\rm SFR$ relation of \citet{kennicutt98},
assuming a Salpeter IMF. We assume $\beta=1.5$ and a dust temperature
of $T=40~\rm K$ for the modified blackbody, similar to what was done
by \citet{wu12} to fit ground-based submm observations of Hot
DOGs. This temperature is chosen as it is representative of the
hottest dust emission associated with star formation
\citep{magnelli12}; lower temperatures will result in lower $\rm SFR$
estimates. The best-fit modified black-body, shown by the solid gray
line in Figure \ref{fg:sed}, yields $\rm SFR_{\rm IR}^{\rm
  Lim}=2500\pm 500~M_{\odot}~\rm yr^{-1}$ if we assume the limit case
where all luminosity of this component is powered by
star-formation. Using the 90\% confidence level, shown by the dotted
gray line in Figure \ref{fg:sed}, we can place the conservative limit
of $\rm SFR_{\rm IR} < 3350~M_{\odot}~\rm yr^{-1}$.

Considering the lower limit derived earlier from the optical SED of
$\rm SFR_{\rm Min} \gtrsim 1000~M_{\odot}~\rm yr^{-1}$, we conclude it
is possible that the UV/optical SED is powered by
star-formation. However, we consider it less likely than the
reflection scenario because a) the SED fit is much poorer than with
our 2AGN model, and b) as discussed earlier, it would be a level of
unobscured star-formation that has not been observed before
\citep{barger14}. If the UV/optical SED is dominated by
star-formation, the upper bound on the stellar mass of $M_*^{\rm Max}
= 7\times 10^{11}~M_{\odot}$ combined with the lower bound on SFR of
$10^{3}~M_{\odot}~\rm yr^{-1}$ implies a specific star formation rate
$> 1.4\times 10^{-9}~\rm yr^{-1}$, i.e., that the host galaxy would be
doubling its stellar mass in a timescale shorter than $700~\rm Myr$.

For completeness, we also estimate the maximum star formation rate
allowed by the X-ray data in addition to the primary AGN component. We
used the relation between SFR and integrated 0.5--8~keV X-ray
luminosity outlined in \citet{mineo14}: $L_{\rm X} / \rm SFR = 4\times
10^{39}~(\rm erg~\rm s^{-1})/(M_{\odot}~\rm yr^{-1})$. These authors
found the total 0.5--8~keV X-ray emission from star formation was
$\sim$2/3 from X-ray binaries (XRBs), and $\sim$1/3 from diffuse
plasma emission from the ISM. For the plasma emission, we adopted a
Mekal plasma model \citep{mewe85} with a rest-frame temperature set to
0.25~keV \citep[the average found by][]{mineo12}. Additionally, at the
star formation rates relevant here, we assumed that the integrated
spectrum from the XRB would be dominated by ultraluminous X-ray
sources \citep[ULXs;][for a recent review]{feng11}, and adopted a
simple model based on recent {\it{NuSTAR}} observations of ULXs
\citep[][]{bachetti13,walton13,walton14,walton15a,walton15b,rana15,mukherjee15},
approximating the spectrum with a cutoff powerlaw model with $\Gamma =
1.5$ and $E_{\rm cut} = 7~\rm keV$, scaled to $z=2.1$. Both components
are then modified by the neutral absorption column of $1.3\times
10^{22}~\rm cm^{-2}$ indicated by the optical emission (also assumed
to be at $z=2.1$). Setting the XRB and diffuse plasma contributions to
2/3 and 1/3 of the total SFR X-ray emission in the rest-frame
0.5--8~keV bandpass, respectively, and allowing the total emission to
vary, we obtain a 90\% upper limit of $\sim5500~M_{\odot}~\rm
yr^{-1}$, somewhat less constraining than, but consistent with, the
upper limit obtained from the {\it{Herschel}} observations. We also
note for completeness that the star-formation constraints from the
FIRST radio observations are poorer than for the X-rays, as the upper
bound of 1~mJy at observed 1.4 GHz flux implies $\rm SFR_{1.4~\rm
  GHz}<7600~M_{\odot}~\rm yr^{-1}$ according to the relation of
\citet{murphy11} and assuming $F_{\nu}\propto \nu^{-0.2}$.

\section{Conclusions}\label{sec:conclusions}

We have introduced an interesting subsample of Hot DOGs that show
UV/optical broad-band emission significantly in excess of what is
typically expected for star formation. Based on UV through mid-IR
photometry and spectroscopy for a large sample of objects, we find
that BHDs constitute $\sim 8\%$ of Hot DOGs, although this number is
considerably uncertain due to the complex selection function.

We argue that the blue excess can most naturally be explained by three
different scenarios, namely: (i) light leaked from the hyper-luminous,
highly obscured AGN that dominates the IR emission, either by
reflection off dust grains or free electrons, or by an opening between
dust clouds allowing a direct line-of-sight to a fraction of the
accretion disk; (ii) a second, less luminous and largely unobscured
AGN in the system; or (iii) a young massive coeval starburst. While
our current data does not allow us to generally differentiate between
these scenarios in most objects, we argue additional observations can
help disentangle them. In particular we note that the detection of
rest-frame UV/optical variability could confirm the AGN nature of the
blue excess, and that spectropolarimetry could confirm the reflection
scenario. Similarly, high spatial resolution UV/optical imaging could
identify the two nuclei of a dual AGN, or confirm the presence of a
galaxy-wide massive young starburst. We also argue that X-ray
observations would allow us to at least partially differentiate
between these scenarios with some confidence. One of the BHDs found on
this article, namely \sname, was serendipitously imaged in the
{\it{Chandra}}/ACIS-I 174.5~ks observations of the LALA field by
\citet{wang07}, and we use these data to analyze the different
scenarios outlined above. With this depth, \sname\ is the Hot DOG with
the best X-ray coverage to date.

We find that the X-ray spectrum of \sname\ is dominated by the higher
energy emission. Of the 94 photons detected in the 0.3--8~keV energy
range, 80 have energies between 2 and 8~keV. Using the models of
\citet{brightman11}, we find that the X-ray spectrum is well fit by an
absorbed AGN with $\Gamma=1.6^{+0.8}_{-0.0}$, $N_{\rm
  H}=6.3^{+8.1}_{-2.1} \times 10^{23}~\rm cm^{-2}$ and an intrinsic
luminosity of $\log L_{2-10\rm keV} / \rm erg~\rm s^{-1} =
44.9^{+0.86}_{-0.14}$. We show that the values of $N_{\rm H}$ and
$\log L_{2-10\rm keV}$ are consistent with those expected from the IR
properties of the AGN through correlations established in the
literature.

Furthermore, we find that the ACIS-I observations strongly limit the
contribution from a hypothetical secondary, lightly absorbed AGN, and
show that a component consistent with the AGN derived from the
UV/optical SED is ruled out with $>99.9\%$ confidence. Hence, we
conclude that the excess blue emission in \sname\ is highly unlikely
to be contributed by a secondary, less luminous AGN in the system. The
lack of detection of a secondary component in the X-rays is consistent
with the scenario of a single partially-covered AGN, but leaked AGN
light due to partial coverage implies an unlikely dust geometry, and
we consider reflected light from a single AGN to be a more likely
explanation. The X-ray observations are also consistent with an
extreme, coeval starburst. Using observations from
{\it{Herschel}}/SPIRE, we show that the IR emission puts a robust
upper limit of $\rm SFR_{\rm IR}<3350~M_{\odot}~\rm yr^{-1}$, while
the rest-frame UV/optical SED requires $\rm SFR \gtrsim
1000~M_{\odot}~\rm yr^{-1}$ to be powered by the star formation,
showing that it is possible the UV/optical emission is powered by
star-formation. However, we consider this scenario less likely the
than the reflection one because a) the fit to the optical data is much
worse than for the AGN model, and b) the required unobscured SFR would
be much larger than the highest observed in Lyman-break galaxies
\citep{barger14}. Hence, the reflection scenario either by dust grains
or free electrons is the most likely one to explain the nature of the
blue excess in \sname, although key details of the model and an
in-depth analysis must be done to fully ascertain its
likelihood. Further testing and constraints of this scenario can be
obtained through deep spectropolarimetric observations to determine
the polarization fraction of the UV/optical emission. Additionally,
high-spatial resolution UV/optical imaging can offer insight into
whether there is a strong galaxy-wide star-burst or if the excess blue
emission is concentrated in the nucleus, as would be expected for the
reflection scenario. A recently approved {\it{Chandra}}/{\it{HST}}
observing program (PI: Assef, Proposal ID: 17700696) will obtain X-ray
and multi-wavelength UV/optical imaging observations for two
additional BHDs, as well as UV/optical imaging for \sname, allowing us
to probe this. These two additional BHDs are also being observed as
part of an ALMA program aimed at studying the \cii\ and far-IR
continuum of Hot DOGs (PI: Assef, Proposal ID:2013.1.00576.S and
2015.1.00612.S). The first results of this program are reported by
\citet{diaz15}. Additional, high spatial-resolution ALMA observations
of the CO emission lines and longer wavelength far-IR continuum in
these objects would determine the extension of the possible
starburst. The combination of these observations will further probe
the nature of these intriguing objects.

\acknowledgments

We thank Sangeeta Malhotra and James Rhoads for providing us with
optical imaging data for the Cetus field. We also thank the anonymous
referee for useful comments and suggestions. RJA was supported by
Gemini-CONICYT grant number 32120009 and FONDECYT grant number
1151408. FEB acknowledges support from CONICYT-Chile (Basal-CATA
PFB-06/2007, FONDECYT 1141218, ``EMBIGGEN'' Anillo ACT1101), and the
Ministry of Economy, Development, and Tourism's Millennium Science
Initiative through grant IC120009, awarded to The Millennium Institute
of Astrophysics, MAS. T.D-S. acknowledges support from ALMA-CONYCIT
project 31130005 and FONDECYT 1151239. This material is based upon
work supported by the National Aeronautics and Space Administration
under Proposal No. 13-ADAP13-0092 issued through the Astrophysics Data
Analysis Program. The scientific results reported in this article are
based to a significant degree on data obtained from the Chandra Data
Archive. This publication makes use of data products from the
Wide-field Infrared Survey Explorer, which is a joint project of the
University of California, Los Angeles, and the Jet Propulsion
Laboratory/California Institute of Technology, funded by the National
Aeronautics and Space Administration. This work is based in part on
observations made with the {\it{Spitzer Space Telescope}}, which is
operated by the Jet Propulsion Laboratory, California Institute of
Technology under a contract with NASA. The WIYN Observatory is a joint
facility of the University of Wisconsin-Madison, Indiana University,
Yale University, and the National Optical Astronomy
Observatory. Funding for SDSS-III has been provided by the Alfred
P. Sloan Foundation, the Participating Institutions, the National
Science Foundation, and the U.S. Department of Energy Office of
Science. The SDSS-III web site is
\url{http://www.sdss3.org/}. SDSS-III is managed by the Astrophysical
Research Consortium for the Participating Institutions of the SDSS-III
Collaboration including the University of Arizona, the Brazilian
Participation Group, Brookhaven National Laboratory, Carnegie Mellon
University, University of Florida, the French Participation Group, the
German Participation Group, Harvard University, the Instituto de
Astrofisica de Canarias, the Michigan State/Notre Dame/JINA
Participation Group, Johns Hopkins University, Lawrence Berkeley
National Laboratory, Max Planck Institute for Astrophysics, Max Planck
Institute for Extraterrestrial Physics, New Mexico State University,
New York University, Ohio State University, Pennsylvania State
University, University of Portsmouth, Princeton University, the
Spanish Participation Group, University of Tokyo, University of Utah,
Vanderbilt University, University of Virginia, University of
Washington, and Yale University. Some of the observations reported
here were obtained at the MMT Observatory, a joint facility of the
Smithsonian Institution and the University of Arizona.

\end{document}